\documentstyle[prb,aps,epsf,floats,goodfloat]{revtex}

\newcommand{\bc}{\begin{center}}
\newcommand{\ec}{\end{center}}
\newcommand{\be}{\begin{equation}}
\newcommand{\ee}{\end{equation}}
\newcommand{\beqn}{\begin{eqnarray}}
\newcommand{\eeqn}{\end{eqnarray}}
\begin{document}
\draft

\twocolumn[\hsize\textwidth\columnwidth\hsize\csname@twocolumnfalse%
\endcsname

\title{
Evidence for a Trivial Ground State Structure in the Two-Dimensional Ising Spin
Glass
}

\author{Matteo Palassini and A. P. Young}
\address{Department of Physics, University of California, Santa Cruz, 
CA 95064}

\date{\today}

\maketitle

\begin{abstract}
We study how the ground state of the two-dimensional Ising spin glass with
Gaussian interactions in zero magnetic field changes on altering the boundary
conditions.  The probability that relative spin orientations change in a region
far from the boundary goes to zero with the (linear) size of the system $L$
like $L^{-\lambda}$, where $\lambda = -0.70 \pm 0.08$. We argue
that $\lambda$ is equal to $d-d_f$ where $d\ (=2)$ is the dimension of the
system and $d_f$ is the fractal dimension of a domain wall induced by changes
in the boundary conditions. Our value for $d_f$ is consistent with 
earlier estimates. These results show that, at zero
temperature, there is only
a single pure state (plus the state with all spins flipped)
in agreement with the predictions of the droplet model.

\end{abstract}

]

The nature of the ordering in spin glasses below the transition temperature,
$T_c$, remains rather poorly understood. For the infinite
range model, the replica symmetry breaking solution of
Parisi\cite{parisi,mpv,by} is generally believed to be correct. An important
aspect of this solution is that the order parameter is a non-trivial
distribution, $P(q)$, where $q$ describes the overlap of the spin configuration
between two copies of the system with identical interactions. The distribution
is non-trivial because very different spin configurations occur with
significant statistical weight.  One loosely says that the system can be in
many ``pure states''.  Monte Carlo simulations on (more realistic)
short range models on quite small lattices\cite{rby,marinari},
find a non-trivial $P(q)$ with a
weight at $q=0$ which is independent of system size (for the range of sizes
studied), as predicted by the Parisi theory.

An alternative approach, the ``droplet model'', has been proposed by Fisher and
Huse\cite{fh} (see also Refs.~\onlinecite{bm,mcmillan}).
Thermodynamic states and pure states are defined
precisely by considering correlation
functions of spins in a region small compared with the system size and
far from the boundary, and asking whether they change or not
upon changing the boundary conditions as the (linear)
system size $L$, tends to
infinity. Each different set of correlation functions corresponds to a
different thermodynamic state.
The droplet theory, the Parisi theory, and some other scenarios have been
studied in detail by Newman and Stein\cite{ns-old,ns-new}.

By making some plausible and self-consistent assumptions, the droplet theory
predicts that the structure of pure states is trivial in short range spin
glasses below $T_c$. In zero field\cite{h=0}, trivial pure state structure
means that any thermodynamic state is a combination of just two distinct pure
states, related by flipping all the spins, which have the same
free energy by symmetry.
If one looks at the whole system, rather than a relatively small region far
from the boundary, one might note that part of the system is in one
pure state and the other part in the spin-flipped state, with a
domain wall between them.
Hence a global quantity like $P(q)$ could
have a non-trivial form\cite{domain} even though the structure of pure
states is actually trivial\cite{pq_trivial}.

To unambiguously distinguish between the droplet and Parisi pictures it is
therefore better to study correlation functions, such as the
overlap distribution, in
a finite region\cite{ns-new} far from the boundary, since the probability that
the domain wall goes through this region vanishes as $L \to\infty$.
More precisely, one should
investigate whether these correlation functions
change when the boundary conditions are
changed.  To our knowledge, however, this has not
been done before\cite{marinari-block}. 

Here, we perform such calculations numerically for the
ground states of the Ising spin glass with Gaussian
interactions in two dimensions.
Although there is no spin glass order at finite temperature in this system,
there {\em is}\/ (complete) spin glass order in the {\em ground state}\/, so
one can investigate the question of the number of pure states {\em at zero
temperature}\/\cite{not-pq}.  Two dimensions has the additional advantage that
there are efficient algorithms for computing exact ground
states\cite{rieger-review,rieger}
and so quite large sizes can be investigated.  We find that the probability for
the spin configuration in the center to change, when the boundary conditions
are altered, goes to zero like $L^{-\lambda}$ as $L$ increases, where
$\lambda$ can be related to the fractal dimension of a domain wall which is
induced by the boundary condition change.  This result shows that there
is only a single pure state at $T=0$ ({\em i.e.}\/ a single ground state), plus
the state with all spins flipped, in agreement with the droplet theory.

The Hamiltonian is given by
\begin{equation}
{\cal H} = -\sum_{\langle i,j \rangle} J_{ij} S_i S_j ,
\label{ham}
\end{equation}
where the sites $i$ lie on the sites of an $L\times L$ square lattice with
$L \le 30$, $S_i=\pm
1$, and the $J_{ij}$ are nearest-neighbor interactions chosen according to a
Gaussian distribution with zero mean and standard deviation unity. Initially,
we impose periodic boundary conditions, denoted by ``P''. Since the
distribution of the interactions, $J_{ij}$, is continuous, the ground state is
unique (apart from the equivalent state obtained by flipping all the spins).
We determine the energy and spin configuration of the ground state for a given
set of bonds. Next we impose anti-periodic conditions (``AP'') along {\em
one}\/ direction, which is completely equivalent to changing the sign of the
interactions along this boundary, and recompute the ground state. Finally we
change the sign of  half the bonds at random along this boundary, which we
denote by ``R''.  Note that the different boundary conditions correspond to
different choices of the interactions which occur with the same probability.
Hence they are statistically equivalent.

For the smaller sizes, $L\le 8$, we compute the ground state by rapidly
quenching from a randomly chosen spin configuration, and repeating many times
until we are confident that the ground state energy has been found. For the two
largest sizes, $L=16$ and 30, this is impractical so instead we use the Cologne
spin glass ground state server\cite{sg-server}. We repeat the calculation of
the ground state for the three copies with different boundary conditions for a
minimum of 2000 samples for each size.

Next we discuss how to study the dependence of the spin configuration on
boundary conditions. We consider a central block containing $N_B = L_B^2$
spins, and ask
if the correlation functions between two spins, $i$ and $j$ say, in the block
depend on the boundary conditions, {\em i.e.}\/ whether $\langle S_i^\alpha
S_j^\alpha
\rangle_T - \langle S_i^\beta S_j^\beta \rangle_T$
is non zero for $L\to \infty$,
where $\alpha$ and $\beta$ refer to two distinct boundary conditions, P, AP, or
R here, $S_i^\alpha$ refers to a spin in the copy with the $\alpha$ boundary
condition, and $\langle \cdots \rangle_T$ denotes a thermal
average. We consider even spin correlation functions because our boundary
conditions do not distinguish between states which differ by flipping all the
spins. 
Since the difference can have either sign, it is convenient to 
consider its square, $( \langle S_i^\alpha S_j^\alpha \rangle_T
 - \langle S_i^\beta S_j^\beta \rangle_T)^2$.
If we sum over all the spins in the block, normalize, and average over
disorder, it is easy to see
that this becomes 
\begin{equation}
\Delta = 
\left\langle \left(q^B_{\alpha\alpha}\right)^2 +
\left(q^B_{\beta\beta}\right)^2 -
2 \left(q^B_{\alpha\beta}\right)^2 \right\rangle ,
\label{delta}
\end{equation}
where 
\begin{equation}
q^B_{\alpha\beta} = {1\over N_B} \sum_{i=1}^{N_B} S_i^\alpha S_i^\beta 
\end{equation}
is the overlap between the block configurations with
$\alpha$ and $\beta$ boundary conditions,
and the brackets $\langle \cdots \rangle$ refer to both a thermal average and
an average over the disorder. 
Eq.~(\ref{delta}) can be written as
\begin{equation}
\Delta = 
\int_{-1}^1 q^2 \left[ P^B_{\alpha\alpha}(q) + P^B_{\beta\beta}(q) - 2
P^B_{\alpha\beta}(q) \right] \, dq ,
\label{delta_pq}
\end{equation}
where
\begin{equation}
P^B_{\alpha\beta}(q) = \left\langle \delta \left( q - q^B_{\alpha\beta}
\right) \right\rangle
\end{equation}
is the
probability distribution for the block overlaps. We have written these
expressions in a general form, valid for $T >0$ as well as $T = 0$.
Similar arguments can be made for correlations of a larger
number of spins, which leads to expressions like Eq.~(\ref{delta_pq}) but with 
higher moments of the overlap distributions.
Hence the crucial quantity is the
difference in the 
block spin overlap distributions with different boundary conditions
which occurs in Eq.~(\ref{delta_pq}),
{\em i.e.}\/
\begin{equation}
\Delta P^B_{\alpha\beta}(q) \equiv
P^B_{\alpha\alpha}(q) + P^B_{\beta\beta}(q) - 2 P^B_{\alpha\beta}(q) .
\end{equation}
If this difference 
tends to zero as $L\to\infty$ then the droplet picture is valid. 
We emphasize that this test
does {\em not} require the size of the block to also become large.

Specializing now to $T=0$,
$P^B_{\alpha\alpha}(q)$ is just the sum of two delta functions
with equal weight at $q=\pm 1$,
since the ground state is unique (apart from overall spin reversal).
Hence, at $T=0$, it is sufficient to investigate
the block overlap distribution $P^B_{\alpha\beta}(q)$
with $\alpha \ne \beta$. We
calculate this for $\alpha = $ P, and $\beta = $ AP and R.

\begin{figure}
\epsfxsize=\columnwidth\epsfbox{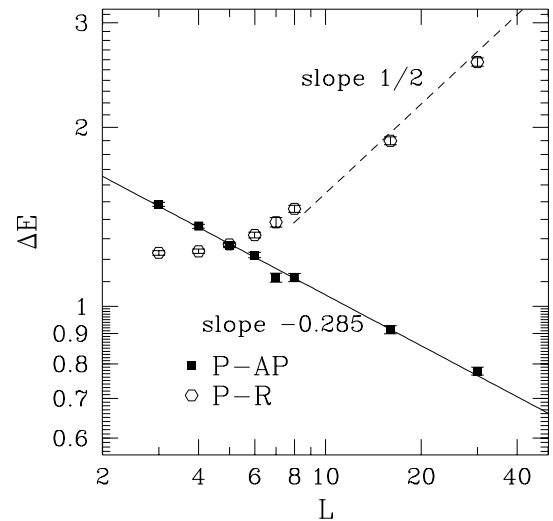}
\caption{
A plot of the root mean square ground state energy differences
$\Delta E_{P,AP}$
and $\Delta E_{P,R}$ for different sizes up to $L=30$.
}
\label{de2}
\end{figure}

Now we discuss our results,
for which we take $L_B=2$. 
First of all, Fig.~\ref{de2} shows data for the
root mean square difference in ground state energy,
\begin{equation}
\Delta E_{\alpha\beta} \equiv
\langle (  E^0_\alpha - E^0_\beta)^2 \rangle^{1/2}
\end{equation}
with $E^0$ the total ground state energy (not the energy per spin),
for $\alpha = $ P and $\beta = $ AP and R. One sees that
$\Delta E_{P,AP}$ goes
to zero like $L^{-\theta}$ as $L$ increases, where $\theta = 0.285 \pm 0.020$.
This is in agreement with earlier
work\cite{rieger,other-theta}. The negative value means that large domains cost
very little energy and so the order in the ground state will spontaneously
break up at any finite temperature, showing that $T_c=0$. 

The results for $\Delta E_{P,R}$
are quite different, however, {\em increasing}
with $L$, roughly as $L^{1/2}$ for large $L$, rather than decreasing. This
difference is easily understood, since the defect (i.e. the region where the
energy is locally different for the two boundary conditions) can be
{\em locally} removed in the P-AP case, by changing the sign of the spins to
one side of the boundary. The defect will then be a single domain wall
somewhere in the sample not necessarily near the boundary. However, this
can not be done for the P-R case and a part of the defect, with an energy which
one could guess to be $L^{(d-1)/2}$ in $d$-dimensions, will stay close to the
boundary, in addition to a domain wall which could be arbitrarily far away.

\begin{figure}
\epsfxsize=\columnwidth\epsfbox{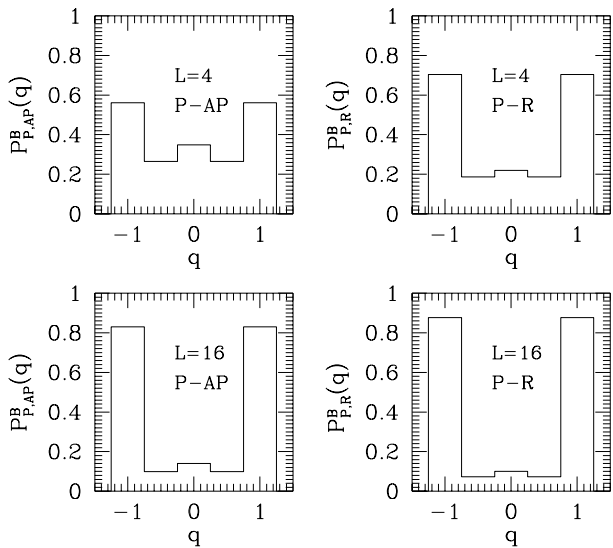}
\caption{
A plot of the block overlaps, $P^B_{P,AP}(q)$ and $P^B_{P,R}(q)$, for $L=4$ and
16, with block size $L_B=2$.
Note that the allowed values of
$q$ are $0, \pm 0.5$ and $\pm 1$.
The left hand column is for the P-AP overlap and the right hand column for
the P-R overlap. The top row is for $L=4$ and the bottom row for $L=16$.  The
data is normalized so that the area under the histograms is unity.
}
\label{hist_multi}
\end{figure}

We show some of our data for the block overlaps in Fig.~\ref{hist_multi}. The
results for the P-AP and P-R overlaps are qualitatively similar to each other,
with the weight away from the peaks at $q=\pm 1$ dropping as $L$ increases.

\begin{figure}
\epsfxsize=\columnwidth\epsfbox{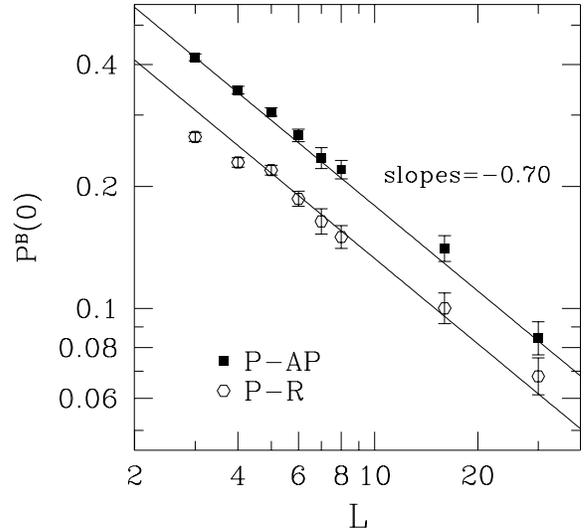}
\caption{
A plot of the block overlaps at $q=0$, $P^B_{P,AP}(0)$
and $P^B_{P,R}(0)$, for different sizes up to $L=30$ with block size $L_B$
equal to 2.
}
\label{p0}
\end{figure}

We characterize this trend by the
weight at $q=0$ and show the results in Fig.~\ref{p0}. For both anti-periodic
and random boundary condition changes, the weight at $q=0$ vanishes like
$L^{-\lambda}$, where $\lambda = -0.70 \pm 0.08$. This value is
easy to understand since $P^B_{\alpha\ne\beta}(0)$ is just the probability
that the domain wall
bisects the block. If the fractal dimension of the domain wall is $d_f$
then, generalizing to $d$-dimensions, the probability that it goes through any
small region is proportional to $L^{-(d-d_f)}$. This immediately gives
$d_f = 1.30 \pm0.08$, which is consistent with
other estimates for $d_f$: $1.26\pm 0.03$ by
Bray and Moore\cite{other-df}, $1.34\pm 0.10$
by Rieger et al.\cite{rieger} and $1.31 \pm 0.1$ (for a related
model) by Gingras\cite{gingras}.

Earlier calculations have investigated some effects of changing the boundary
conditions from periodic to anti-periodic, usually just the change in the
ground state energy, though Bray and Moore\cite{other-df}
and Rieger\cite{rieger} have also calculated
the fractal dimension of the domain wall.
They obtain a value less than $d$ (as
noted above), which implies that $P^B_{P,AP}(0)$ vanishes for large $L$, as we
find explicitly here.
However, as noted in our discussion of Fig.~\ref{de2}, anti-periodic boundary
conditions are special since the defect can be locally removed by flipping the
spins to one side of it. This is why we also investigate random boundary
condition changes, for which the defect cannot be locally eliminated.  An
important result of our work is that the fractal dimension of the domain wall
is the same in both cases.

For each configuration of the bonds in the bulk we have only studied a single
random change in the boundary conditions. It would be interesting to get
statistics on a large number of boundary condition changes to see if the
probability for the domain wall to go through the central block obtained by
averaging over boundary conditions for a single large sample is the same as we
find here by averaging over samples.  It would also be interesting to
investigate boundary conditions which are optimized to minimize the ground
state energy, i.e. which are correlated with the bonds in the bulk.


Recently, we have been able to
perform calculations similar to those presented here for the
three-dimensional spin glass\cite{py3d}, which has a finite $T_c$.
There too we find evidence for
a unique ground state.

After this work was virtually complete, we became aware of related work by
Middleton\cite{middleton}.
Whereas we start with {\em periodic} boundary
conditions
Middleton takes {\em free} boundary conditions which allows him to
find a ground state (for the models studied) in polynomial
time, permitting the study of very large sizes,
up to $L=512$.
In this approach one can only perturb the system far away from the central
region by making it grow bigger. Hence there are
{\em three} relevant
lengths: the block size, which we call
$L^B$, the size inside which the bonds
are not changed (which we will call $L_{mid}$), and the overall size $L$. One
needs $L_{mid} \gg L^B$ and at least some data which also satisfies $L \gg
L_{mid}$.  Hence the largest sizes $L$ that Middleton studies need to be
{\em very} large. 
In our work, we only have one inequality to satisfy, $L \gg L^B$, rather than
two, so the sizes do not need to be as large.
Overall, the two approaches are complementary and, in our view, have
similar validity for the two-dimensional spin glass. However we believe that
our approach is preferable for the
three-dimensional spin glass, for which there are no polynomial algorithms,
since it requires smaller sizes.

This work was supported by the National Science Foundation under grant DMR
9713977.  M.P. is supported by University of California, EAP Program, and by a
fellowship of Fondazione Angelo Della Riccia. We would like to thank
D.~L.~Stein and G.~Parisi for their comments on an earlier version of the
manuscript.  We would also like to thank Prof. M.~J\"unger and his group at the
University of Cologne for putting their spin glass ground state server in the
public domain.


\begin{references}
\bibitem{h=0}
    We only consider the case of zero field in this paper.
\bibitem{pq_trivial}
    It is also possible that $P(q)$ could be trivial, even though the
    structure of pure states is non-trivial, because $P(q)$ detects only
    states which differ in free energy by a finite amount from the state of
    lowest free energy, whereas changing the boundary conditions can alter the
    free energy of states by an amount which diverges with the size of the
    system, though less than extensively. Hence it is possible that $P(q)$
    could miss some pure states.
\bibitem{not-pq}
    Note that $P(q)$ is not a useful quantity at $T=0$ because both the Parisi
    and droplet pictures predict two delta functions with equal weight at
    $q=\pm 1$ in a finite system. This is a trivial consequence of the ground
    state being only doubly degenerate (the two states being related by time
    reversal) when the bond distribution is continuous.
\end{references}
\end{document}